\documentclass[journal=jacsat,manuscript=article]{achemso}
\setkeys{acs}{etalmode=truncate,maxauthors=10}
\usepackage[version=3]{mhchem} % Formula subscripts using \ce{}
\usepackage{xcolor}
\usepackage{soul}

\usepackage{amsmath}
\usepackage{amssymb}
\usepackage{bm}           % Bold math w/ amsmath
\usepackage{mathrsfs}
\usepackage{booktabs,enumitem} % For NiceTabular environment
\usepackage[footnotehyper]{nicematrix} % For NiceTabular environment
\usepackage{multirow}
\usepackage{float}
\usepackage{etoolbox}  % Allows patching \email command
\usepackage[labelfont=bf, skip=0pt]{caption} %font=footnotesize,

\usepackage{comment}

\makeatletter
\patchcmd{\acs@contact@details}{E-mail}{*\,Email}{}{}  % patch email command to add *
\makeatother

\author{Justin Airas}
\author{Bin Zhang}
\email{binz@mit.edu}
\affiliation{Department of Chemistry, Massachusetts Institute of Technology, Cambridge, MA, USA}

\title[An \textsf{achemso} demo]
  {Knowledge Distillation of a Protein Language Model Yields a Foundational Implicit Solvent Model}

\begin{document}

% especially powerful in the field of chemistry. 

\begin{abstract}
Implicit solvent models (ISMs) promise to deliver the accuracy of explicit solvent simulations at a fraction of the computational cost. However, despite decades of development, their accuracy has remained insufficient for many critical applications, particularly for simulating protein folding and the behavior of intrinsically disordered proteins. Developing a transferable, data-driven ISM that overcomes the limitations of traditional analytical formulas remains a central challenge in computational chemistry. Here we address this challenge by introducing a novel strategy that distills the evolutionary information learned by a protein language model, ESM3, into a computationally efficient graph neural network (GNN). We show that this GNN potential, trained on effective energies from ESM3, is robust enough to drive stable, long-timescale molecular dynamics simulations. When combined with a standard electrostatics term, our hybrid model accurately reproduces protein folding free-energy landscapes and predicts the structural ensembles of intrinsically disordered proteins. This approach yields a single, unified model that is transferable across both folded and disordered protein states, resolving a long-standing limitation of conventional ISMs. By successfully distilling evolutionary knowledge into a physical potential, our work delivers a foundational implicit solvent model poised to accelerate the development of predictive, large-scale simulation tools.
\end{abstract}

%\clearpage
%\newpage
\section{Synopsis}
We demonstrate that a graph neural network can act as a foundational protein implicit solvent model when trained to match the distribution of secondary structure predictions from a protein language model.

\section{TOC Graphic}

\begin{figure}[H]
    \centering
    \includegraphics[]{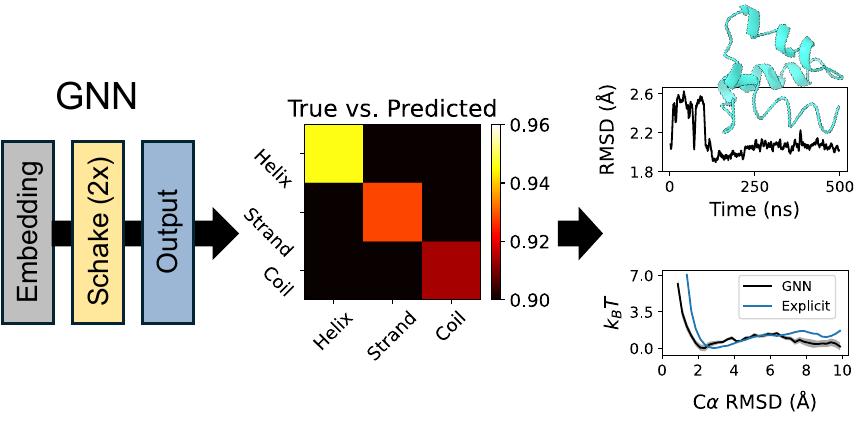}
    \label{TOC}
\end{figure}

\clearpage
\newpage
\section{Introduction}

Implicit solvent models (ISMs) represent a promising middle ground in molecular simulation, offering a significant computational advantage over all-atom explicit solvent simulations while providing greater physical detail than lower-resolution coarse-grained models \cite{latham_improving_2019,latham_maximum_2020,liu_openabc_2023,liu_toward_2025,chen_associative_2018,joseph_physics-driven_2021,dignon_sequence_2018,baul_sequence_2019}. The middle-ground promise of ISMs has generated significant enthusiasm. Nevertheless, despite decades of refinement~\cite{hawkins_parametrized_1996,onufriev_exploring_2004,mongan_generalized_2007,nguyen_improved_2013,gallicchio_agbnp_2004,chen_balancing_2006,lee_novel_2002,lee_new_2003,im_generalized_2003,haberthur_facts_2008,vitalis_absinth_2009,chekmarev_long-time_2004,lee_optimization_2017,onufriev_generalized_2019,onufriev_water_2018,akkermans_solvation_2017}, their accuracy remains well below that of their explicit solvent counterparts. Traditional ISMs, such as popular Generalized Born (GB) and surface-area-based formulations, systematically misrepresent the balance between solvation and intramolecular interactions, frequently leading to artifacts like the over-compaction of disordered proteins, overstabilization of $\alpha$-helical conformations, and exaggerated protein–protein association energies~\cite{arsiccio_new_2022,greener_differentiable_2024,shao_assessing_2018,lang_generalized_2022,shell_test_2008,best_computational_2017,bottaro_variational_2013}.

These deficiencies stem from two overarching limitations. First, ISMs rely on approximate analytical expressions to compute the solvation free energy, $E_\mathrm{solv}$; such approximations cannot fully capture the complex dependence of $E_\mathrm{solv}$ on molecular composition, geometry, and conformational state. Second, the parameters of ISMs are rarely optimized in a systematic or data-driven manner to reproduce results from experiment or explicit solvent simulations across diverse protein families. The rise of machine learning (ML) offers a path to overcome these challenges. By replacing the approximate analytical formulas with flexible neural network potentials, ML-based force fields could lead to ISMs with much-improved accuracy. Many promising studies have demonstrated the potential of using ML methodologies for force field development.\cite{airas_transferable_2023,katzberger_implicit_2023,katzberger_general_2024,katzberger_rapid_2025,katzberger_transferring_2025,chen_machine_2021,husic_coarse_2020,kozinsky_scaling_2023,wang_improving_2023,brunken_machine_2024,yao_machine_2023,wang_machine_2019,wang_multi-body_2021,fu_forces_2023,durumeric_learning_2024,arts_two_2023,kohler_flow-matching_2023,loose_coarse-graining_2023,durumeric_adversarial-residual-coarse-graining_2019,durumeric_machine_2023,ding_contrastive_2022,ding_optimizing_2024,chmiela_towards_2018,chmiela_accurate_2020,duschatko_thermodynamically_2024,greener_differentiable_2024,yang_slicing_2023,corso_graph_2024,batzner_advancing_2023,eastman_openmm_2024,zheng_predicting_2024,sahrmann_utilizing_2023,bonneau_peering_2025,anstine_machine_2023,cheng_developing_2024,faller_density-based_2024,takaba_machine-learned_2024,wang_espalomacharge_2024,wang_end--end_2022,galvelis_nnpmm_2023,sabanes_zariquiey_enhancing_2024,barnett_neural_2024,yang_construction_2022,charron_navigating_2025,majewski_machine_2023,liao_development_2025,riveros_neat-dna_2025,smith_ani-1_2017,devereux_extending_2020,gao_torchani_2020,unke_biomolecular_2024,zhang_pretraining_2024,klein_operator_2025,plainer_consistent_2025,murtada_md-llm-1_2025}

Graph neural networks\cite{unke_physnet_2019,gasteiger_directional_2022,gasteiger_gemnet_2022,anderson_cormorant_2019,brandstetter_geometric_2022,li_egnn_2022,schutt_equivariant_2021,thomas_tensor_2018,fuchs_se3-transformers_2020,huang_equivariant_2022,schutt_schnet_2018,satorras_en_2022,wang_spatial_2023,tholke_torchmd-net_2022,pelaez_torchmd-net_2024,wang_enhancing_2024,batzner_e3-equivariant_2022,batatia_design_2025,musaelian_learning_2023,batatia_mace_2023,mann_egret-1_2025,airas_scaling_2025,caruso_extending_2025} (GNNs) are particularly well-suited as neural network potentials. Since molecules are naturally structured as graphs, with atoms as nodes and interactions as edges, GNNs provide a native framework for learning complex, geometry-dependent energetic terms. Moreover, their architecture facilitates the straightforward incorporation of fundamental physical principles, such as translational, rotational, and permutational invariance, which is crucial for building robust and generalizable force fields.
However, parameterizing a GNN as an ISM is not trivial. Unlike typical ML force fields trained on quantum mechanical energies, the target $E_\mathrm{solv}$ values are not directly known, precluding simple supervised learning. Several theoretically sound approaches have been introduced to optimize ISMs by matching configurational ensembles from explicit solvent simulations.\cite{chen_machine_2021,katzberger_general_2024,airas_transferable_2023} While powerful, these methods are constrained by the finite set of proteins for which such simulation data is available, raising concerns about the transferability of the resulting models to new and diverse protein systems.

In this study, we introduce a novel training strategy to address this transferability challenge by learning a proxy for $E_\mathrm{solv}$ from a protein language model. We leverage the multimodal model ESM3~\cite{hayes_simulating_2025}, which captures the joint distribution of sequence, structure, and function across billions of proteins and attains near-experimental accuracy in predicting 3D structures directly from sequence. Its conditional probabilities, $P(\text{structure}|\text{sequence})$, and corresponding effective energies, $E=-k_BT\log P$, therefore approximate the true folding free-energy landscape. As solvation dominates folding energetics, these evolution-derived probabilities offer an excellent proxy for solvent-mediated effects.

Putting this hypothesis to the test, we first demonstrate that a GNN with significantly fewer parameters can successfully reproduce the effective energies predicted by ESM3, a remarkable distillation of knowledge. The resulting GNN potential is robust enough to drive long-timescale molecular dynamics (MD) simulations. We show that for several proteins, simulations of up to 500 ns maintain stable structures near their native conformations, providing strong evidence that the evolutionary statistics within ESM3 indeed serve as a high-fidelity proxy for the protein folding free-energy landscape.

To create a physically predictive model, we further combine the distilled GNN potential with a standard GB electrostatic term. This hybrid representation not only reproduces explicit-solvent folding free-energy landscapes with high fidelity but also captures the secondary-structure distributions of intrinsically disordered proteins, an area where traditional ISMs routinely fail. Together, these results demonstrate that our approach yields a genuinely transferable ISM capable of describing both ordered and disordered states within a single framework. By distilling the evolutionary knowledge encoded in a protein language model into a compact neural network potential, we establish the first foundational ISM: a scalable, data-driven model that provides a robust starting point for building the next generation of accurate and efficient protein simulations.

\section{Results}
\subsection{Distilling solvent-sensitive secondary structure preferences into a multiscale GNN}

    To develop an accurate and transferable machine-learning ISM, we require a flexible architecture capable of representing solvent-mediated interactions across diverse proteins. We therefore adopt Schake, a recently introduced multiscale GNN designed for protein-scale systems.\cite{airas_scaling_2025} Schake combines (i) a short-range SAKE message-passing layer \cite{wang_spatial_2023} that encodes detailed chemical interactions with (ii) a long-range SchNet message-passing layer \cite{schutt_schnet_2018} that aggregates coarse-grained structural context, enabling both accuracy and scalability.

    \begin{figure}[t!]
    \centering
    \includegraphics[width=\textwidth]{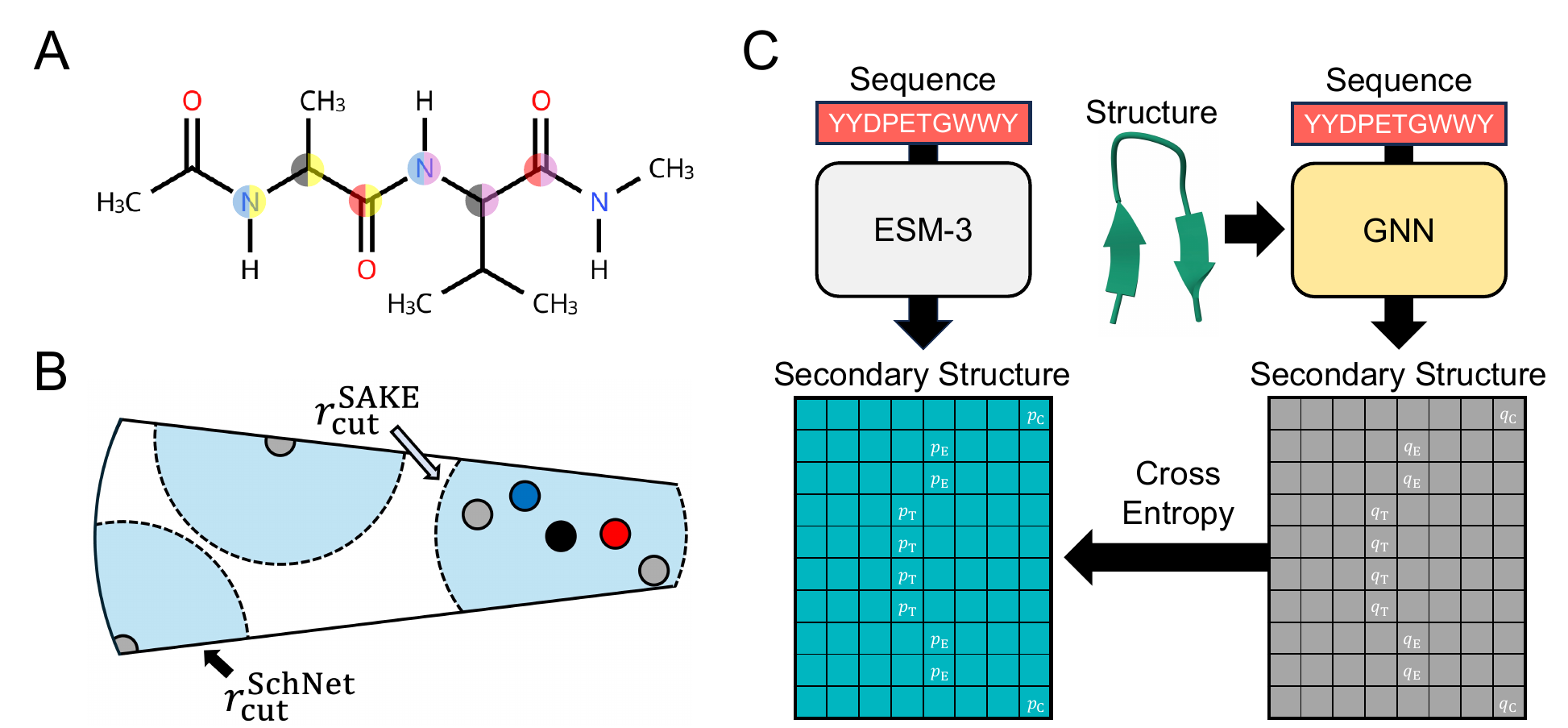}
    \caption{
    \textbf{Knowledge distillation enables training of an efficient, multiscale GNN architecture.}
    (A) Capped Ala-Val dipeptide with the C$\alpha$, C, and N backbone atoms highlighted. These atoms serve as inputs to the GNN. The color of the left half of the circle indicates the backbone atom type, while the color on the right half indicates the amino acid type. 
    (B) Diagram of the Schake multiscale message-passing scheme. The short-ranged SAKE message-passing layer acts on all backbone C$\alpha$, C, and N atoms within the cutoff distance $r_\mathrm{cut}^\mathrm{SAKE} = 1$ nm, while the long-ranged SchNet message-passing layers acts on only C$\alpha$ atoms within $r_\mathrm{cut}^\mathrm{SchNet} = 2.5$ nm.
    (C) Diagram of the knowledge distillation training methodology. By inputting an amino acid sequence, ESM3 is used to predict a matrix of SS8 motif likelihoods. A GNN that inputs both an amino acid sequence and a structure is then trained to match the ESM3-predicted matrix of SS8 motif likelihoods using the cross entropy loss function. The sequence and structure of chignolin CLN025 is shown as an example, with the likelihoods for the folded state motifs shown in the matrices. 
    }
    \label{train_fig}
    \end{figure}

    As noted in the \emph{Introduction}, training ISMs requires either accurate solvation free energies or extensive explicit-solvent configurational ensembles, both of which remain computationally expensive to obtain at the scale needed for broad transferability, although larger datasets are beginning to emerge~\cite{wang_sequence-dependent_2025,vandermeersche_atlas_2024,amaro_need_2025}. We therefore turn to an alternative, data-driven proxy by making use of ESM3, a multimodal protein language model trained across billions of natural protein sequences and structures.\cite{hayes_simulating_2025}

    ESM3 models the joint distribution of protein sequence and structure, enabling residue-level structural predictions from the sequence alone. These predictions reflect the statistical signatures of folding energetics, which are strongly shaped by solvation. Distilling this sequence–structure relationship into Schake would allow the GNN to inherit solvent-sensitive conformational preferences in much the same way that an accurate solvation free energy would.

    While ESM3 produces both secondary- and tertiary-structure distributions at the residue level, we begin by distilling a local observable that is solvent sensitive, interpretable, and naturally compatible with graph-based modeling: %the SS8 secondary-structure motif distribution. 
    the 8 secondary-structure motifs (abbreviated as SS8) detailed by the Define Secondary Structure of Proteins (DSSP) algorithm.\cite{kabsch_dictionary_1983}
    SS8 motifs encode hydrogen-bonding patterns and steric preferences, providing a clean and tractable signal for transferring solvent-dependent structural tendencies from ESM3 into a GNN architecture.

    To transfer the solvent-informed structural knowledge encoded in ESM3, we train Schake to reproduce the ESM3-predicted SS8 motif likelihoods for proteins in the DISPEF-M dataset.\cite{airas_scaling_2025}  Although Schake was originally designed to operate on full atomistic protein structures, we restrict its inputs to the backbone C$\alpha$, C, and N atoms to reduce computational cost while retaining the essential geometric information required for secondary-structure classification. This backbone-only representation allows Schake to capture solvent-sensitive local structural tendencies with minimal overhead (Fig.~\ref{train_fig}A--B). DISPEF-M provides an ideal training set, offering diverse structural folds for $\sim$20,000 proteins together with atomistic structural models.
    
\begin{figure}[t!]
    \centering
    \includegraphics[width=\textwidth]{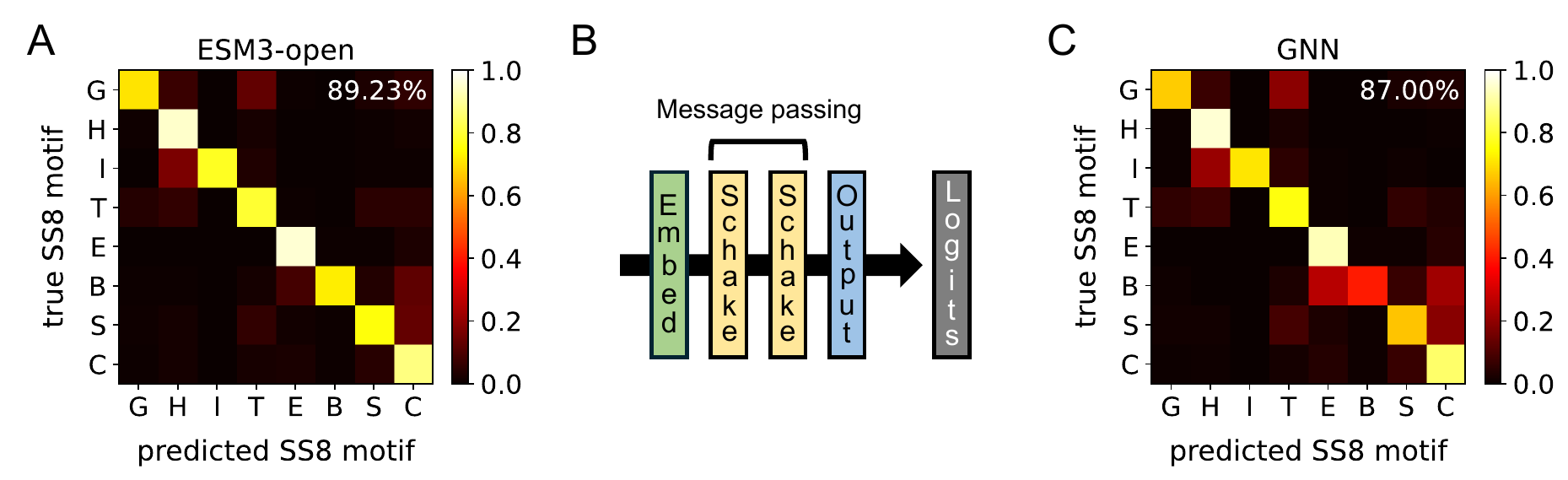}
    \caption{
    \textbf{Schake matches the SS8 motif predictions from ESM3-open.}
    (A) Confusion matrix for ESM3-open SS8 motif predictions. 
    (B) Architecture of the Schake GNN. Backbone atoms are inputted, and SS8 motif logits are outputted. 
    Refer to the \emph{Methods} section for a detailed description of the Schake architecture.
    (C) Confusion matrix for GNN SS8 motif predictions.
    For (A) and (C), the mean predicted likelihood for each SS8 motif conditioned on the true SS8 motif is displayed. The percentage in the upper right corner indicates the mean predicted likelihood of the true SS8 motif (computed by averaging along the diagonal).
    Each letter corresponds to a particular SS8 motif: G to $3_{10}$-helices, H to $\alpha$-helices, I to $\pi$-helices, T to hydrogen-bonded turns, E to $\beta$-sheets, B to $\beta$-bridges, S to non-hydrogen-bonded bends, and C to non-categorized structures.
    }
    \label{confusion_mats}
    \end{figure}

    Despite containing only 45,000 parameters, the distilled Schake model closely matches the SS8 likelihoods of the 1.4 billion-parameter ESM3-open model, achieving an average correct-motif probability of 87.0\% compared to ESM3-open’s 89.2\% (Fig.~\ref{confusion_mats}). 
    Furthermore, measured on a Nvidia L40S GPU over 200 repeated inferences, Schake achieves a mean prediction time of 2.16\,ms for the 80-residue protein $\lambda$-repressor, approximately 9-fold faster than the corresponding ESM3-open inference (mean 19.23\,ms).
    This substantial speedup is crucial for potential downstream integration into molecular simulations.

    To assess the transferability and scalability of the distilled model, we further evaluated Schake on the large-protein subset DISPEF-L. DISPEF-L comprises more than 100,000 proteins ranging from 400 to 800 amino acids, substantially larger than those included in DISPEF-M. Schake maintains strong performance in this much broader regime, achieving an average correct-motif probability of 85.2\% (Fig.\ S1). These results demonstrate that Schake generalizes well beyond its training set and scales effectively to proteins far larger than those encountered during training.
\subsection{GNN-derived energies reliably distinguish folded and unfolded protein states}

In the previous section, we showed that Schake can faithfully reproduce the solvent-sensitive SS8 motif statistics distilled from ESM3, providing a learned signal of how local backbone environments respond to solvation.  These SS8 motif likelihoods naturally induce a configuration-dependent energy, constructed as a sum of local contributions from individual backbone atoms, which we define as
\begin{equation}
E_\mathrm{GNN}^\mathrm{os}(\bm{x}) = -\gamma k_B T
\sum_{i=1}^{3n_\mathrm{res}}
\sum_{j \in S_i}
y_i^{(j)} \log\!\left[q_i^{(j)}(\bm{x})\right],
\tag{1}
\label{Eq:01}
\end{equation}
where $\bm{x}$ denotes the input configuration and $q_i^{(j)}(\bm{x})$ is the GNN-predicted likelihood of SS8 motif $j$ at backbone atom $i$. The index $i$ runs over all backbone atoms (three per residue) used in the GNN predictions. The binary variable $y_i^{(j)}$ indicates which SS8 motifs are present in the reference (folded) structure, such that the resulting one-state (os) energy selectively stabilizes this reference state. The dimensionless scaling factor $\gamma$ controls the overall strength of the GNN-derived energy.

Because the GNN was trained exclusively on folded structures, this energy is expected to behave sensibly for local deviations from the native state: as the conformation perturbs away from the folded structure, the likelihoods assigned to native-state motifs decrease, causing the associated energy to rise.  As a result, $E_\mathrm{GNN}^\mathrm{os}$ is minimized in the vicinity of the native state, stabilizing the folded structure locally.  Whether this behavior persists for large-scale unfolding events, far outside the training distribution, is a key question for determining whether the GNN-derived energy can contribute meaningfully to an ISM.

    To assess this, we next evaluate whether $E_\mathrm{GNN}^\mathrm{os}$ increases appropriately for conformations that depart substantially from the native basin, including globally unfolded and highly extended states.  A meaningful solvation-like correction should assign higher energies to such configurations than to the folded structure, even though the model was trained exclusively on native-like conformations.

    \begin{figure}[t!]
    \centering
    \includegraphics[width=\textwidth]{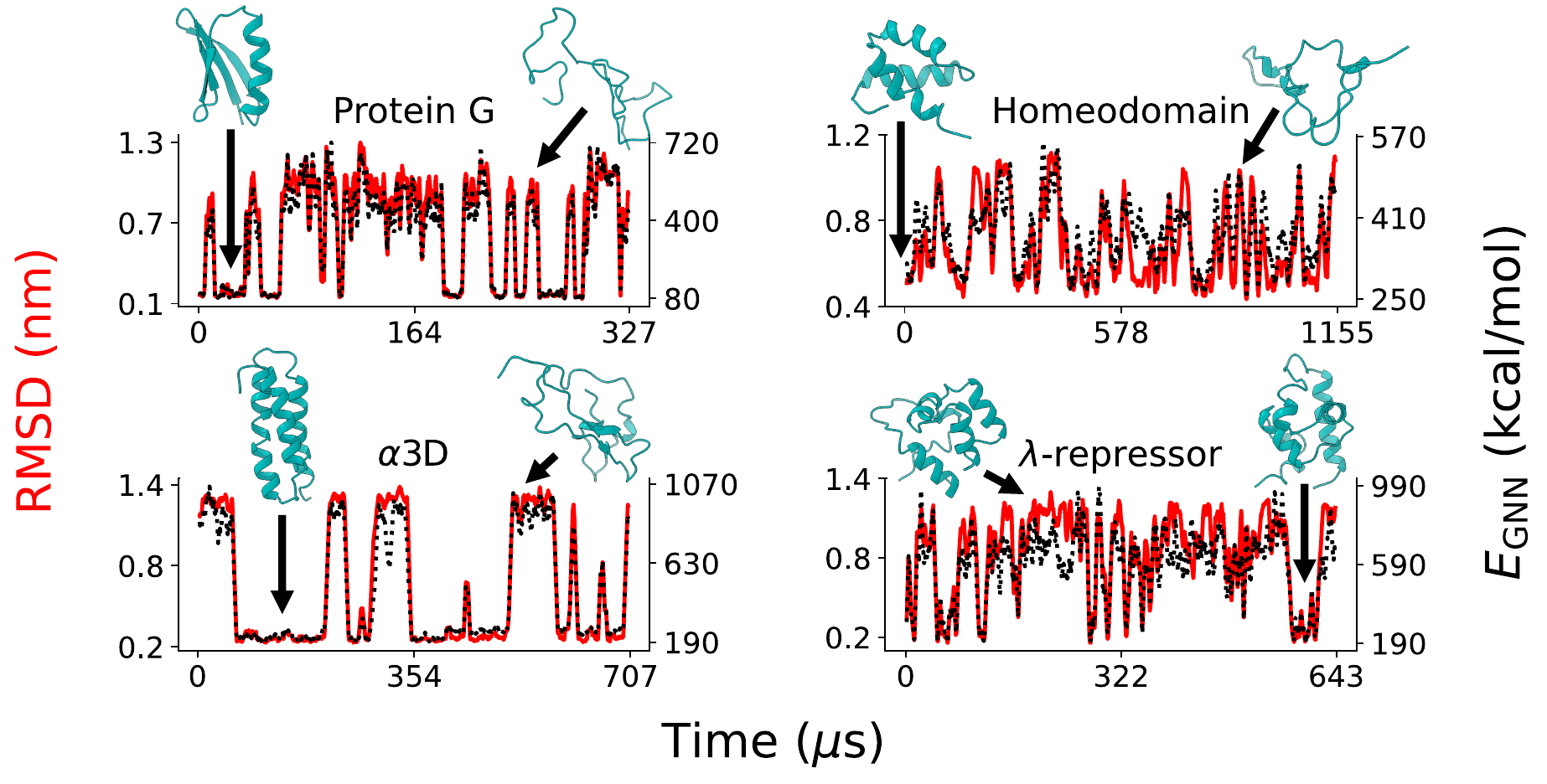}
    \caption{
    \textbf{GNN energy changes closely correspond to structural changes.}
    $E_\mathrm{GNN}^\mathrm{os}$ was computed across long time-scale MD trajectories generated from a previous work.\cite{lindorff-larsen_how_2011} Each trajectory was subset following the procedure used in a previous GNN study.\cite{airas_transferable_2023} Results for the four largest proteins are shown here, while results for the other eight proteins are shown in the Fig.\ S2. RMSD from the folded state (reference structures are detailed in the \emph{Methods} section) is shown in red on the left y-axis, while $E_\mathrm{GNN}^\mathrm{os}$ is shown in black on the right y-axis. Here, the scaling constant $\gamma = 2.5$ and the temperature $T = 300$ K. Note that non-standard amino acids and capping groups 
    are excluded from calculation of $E_\mathrm{GNN}^\mathrm{os}$. For all proteins, the lowest-energy structure is that with the lowest RMSD from the folded state.
    }
    \label{desres_Egnn}
    \end{figure}

We first evaluated the behavior of $E_\mathrm{GNN}^{\mathrm{os}}$ on long atomistic trajectories generated by D.\ E.\ Shaw Research,\cite{lindorff-larsen_how_2011} which contain repeated folding and unfolding events and therefore provide a stringent test of generalization beyond the folded training set.  For these fast-folding proteins, we computed both the RMSD from the native structure and $E_\mathrm{GNN}^{\mathrm{os}}$ as functions of simulation time.  As shown in Fig.~\ref{desres_Egnn} and S2, the RMSD traces display clear two-state switching behavior, with long plateaus in folded or unfolded basins punctuated by rapid transitions between them.  Remarkably, the GNN-derived energy mirrors these transitions almost perfectly: whenever the RMSD increases during an unfolding event, $E_\mathrm{GNN}^{\mathrm{os}}$ rises in tandem, and it returns to a low value immediately upon refolding. Structures closest to the folded state consistently exhibit the lowest $E_\mathrm{GNN}^{\mathrm{os}}$, whereas unfolded conformations yield substantially higher energies. Notably, this behavior is not seen in the GBn2\cite{nguyen_improved_2013} ISM (Fig.\ S3). The tight correlation of $E_\mathrm{GNN}^{\mathrm{os}}$ with RMSD demonstrates that the SS8-based energy assignment remains reliable even under large conformational changes far outside of the training set.

We next tested whether this relationship between unfoldedness and GNN energy persists in much larger proteins and in ensembles deliberately biased toward extended conformations. For each protein, we computed $E_\mathrm{GNN}^{\mathrm{os}}$ directly as a function of the RMSD from the folded structure.  Across all systems examined, $E_\mathrm{GNN}^{\mathrm{os}}$ increases steadily with RMSD (Fig.\ S4): native-like states exhibit the lowest energies, partially unfolded states carry intermediate energies, and highly extended configurations yield the highest energies.  
The consistency of this trend demonstrates that the GNN-derived energy generalizes robustly across sequence length, structural complexity, and sampling regime.

Therefore, the pretrained GNN, without any fine-tuning, produces an energy term that robustly distinguishes folded from unfolded ensembles and penalizes large-scale deviations from the native structure. 
\subsection{GNN-Derived Energies Support Stable ML/MD Simulations}

We next assess whether the learned energetic term, $E_\mathrm{GNN}^{\mathrm{os}}$, can be used directly within MD simulations to propagate large proteins stably over hundreds of nanoseconds.

\begin{figure}[t!]
\centering
\includegraphics[width=0.95\textwidth]{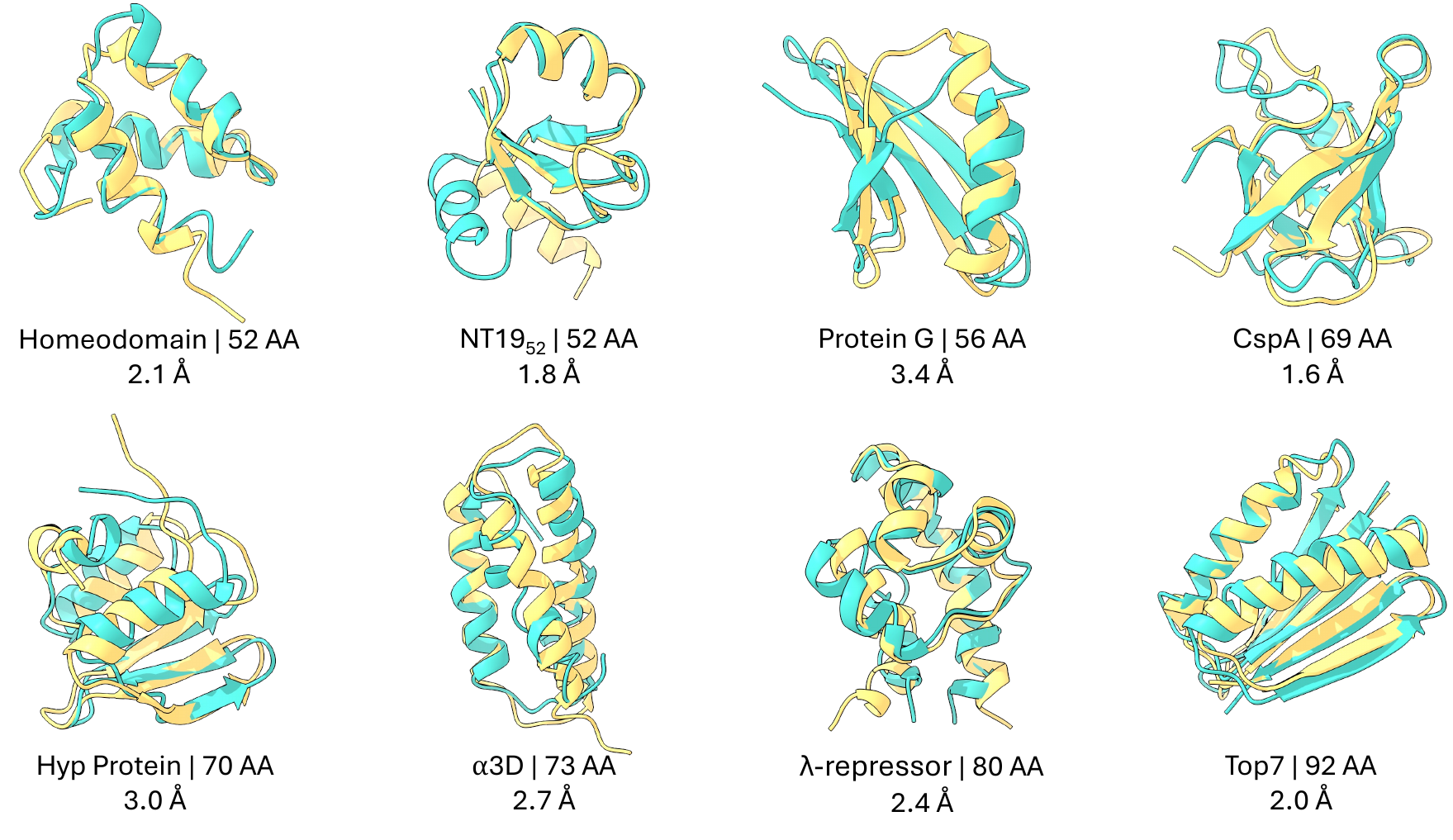}
\caption{
\textbf{ML/MD simulations with Schake maintain structures close to their starting folded conformations.}
Starting folded structures are shown in yellow; ML/MD structures correspond to the lowest-$E_\mathrm{GNN}^{\mathrm{os}}$ conformations matching the median trajectory RMSD and are shown in teal. Each protein is labeled with its name, length, and RMSD from the starting structure. Flexible regions (Fig.~S6, Tab.~S1) are excluded from RMSD calculations.
}
\label{ml_pdb_fig}
\end{figure}

We performed 500 ns of ML/MD simulations for the eight largest proteins examined by \citet{nguyen_folding_2014}, using Schake with $\gamma = 2.5$, beginning from their experimentally determined folded structures. Across this set, Schake consistently preserves native-like conformations (Fig.~\ref{ml_pdb_fig}). 
Even the most divergent configurations sampled remain within 4~\AA{} RMSD of the folded state (Tab.~S2), indicating that the GNN-derived energy not only favors but also dynamically confines each system to the native basin.
Time traces of both RMSD and $E_\mathrm{GNN}^{\mathrm{os}}$ (Fig.~S5) show that energetic fluctuations closely parallel structural deviations, mirroring the tight correspondence between $E_\mathrm{GNN}^{\mathrm{os}}$ and RMSD observed for the D.\ E.~Shaw Research trajectories.

As a control, we performed matched 500 ns GBn2 simulations for the same proteins. In contrast to Schake, three of the GBn2 trajectories sample structures exceeding 4~\AA{} RMSD from the folded state (Fig.~S5, Tab.~S3), consistent with the tendency of classical GB models to overstabilize partially unfolded or misfolded compact configurations. This comparison highlights that the GNN-based energetic term exhibits superior dynamical stability relative to traditional ISMs.

To further examine structural integrity, DSSP analysis was performed for both the ML/MD trajectories and their corresponding folded structures. Five proteins show essentially unchanged secondary-structure content over 500 ns (Tab.~S2). In homeodomain and $\lambda$-repressor, modest reductions in helical content coincide with increases in hydrogen-bonded turn motifs (Fig.~S7), consistent with local rearrangements rather than global unfolding. Protein~G exhibits a local disruption in the $\beta$-sheet between residues~42--46. For comparison, GBn2 trajectories also show deviations in helical content (Tab.~S3), including loss of helical structure in NTL9$_{52}$ that remains stable under Schake.

\subsection{Multi-state energy formulation enables accurate simulation of folded and intrinsically disordered proteins}

The ability to convert SS8 motif likelihoods into an energy function that stabilizes folded conformations in MD simulations is notable in itself. The one-state formulation $E_\mathrm{GNN}^{\text{os}}$ shows that local structural preferences learned by ESM3 can be distilled into a physically meaningful potential capable of identifying and maintaining native basins.

These results naturally raise a deeper question: can such a distilled energy be pushed beyond qualitative stabilization to achieve \emph{quantitative} agreement with explicit solvent simulations? This is far from guaranteed, as ESM3 was not trained on explicit-solvent data and $E_\mathrm{GNN}^{\text{os}}$ is designed to favor the native ensemble rather than balance folded, partially unfolded, and disordered states. Achieving quantitative accuracy therefore requires an energy function that can assign reasonable free energies to alternative local motifs present across unfolded ensembles and IDPs. This motivates the introduction of a \emph{multi-state} energy formulation.

The multi-state (ms) energy evaluates, at each backbone position, the most probable SS8 motif according to the GNN, without privileging any specific reference state:
\begin{equation}
    E_\mathrm{GNN}^{\text{ms}} \left( \bm{x} \right)
    = -\,\gamma k_B T\, \sum_{i=1}^{3n_\mathrm{res}}
    \log \left[\max_{j \in S_i} q_i^{(j)}(\bm{x}) \right].
    \tag{2}
    \label{new_e}
\end{equation}
Notation follows Eq.~\ref{Eq:01}; to maintain differentiability, the $\max$ is smoothly approximated using the LogSumExp function (Eq.~S1). Conceptually, $E_\mathrm{GNN}^{\text{ms}}$ treats the GNN as a local motif evaluator, rewarding any structurally plausible environment and allowing the energy to adapt seamlessly across folded, partially folded, and disordered conformations.

As with $E_\mathrm{GNN}^{\text{os}}$, we computed $E_\mathrm{GNN}^{\text{ms}}$ along the D.\ E.~Shaw Research trajectories.
Across most proteins, $E_\mathrm{GNN}^{\text{ms}}$ correlates strongly with RMSD (Fig.~S8)%; Tab.~S1--S2)
, and for all but the two smallest systems (CLN025 and Trp-cage), the folded state remains the global minimum. 
Thus, the multi-state formulation retains the stabilizing behavior demonstrated in previous sections.

\begin{figure}[t!]
\centering
\includegraphics[width=0.55\textwidth]{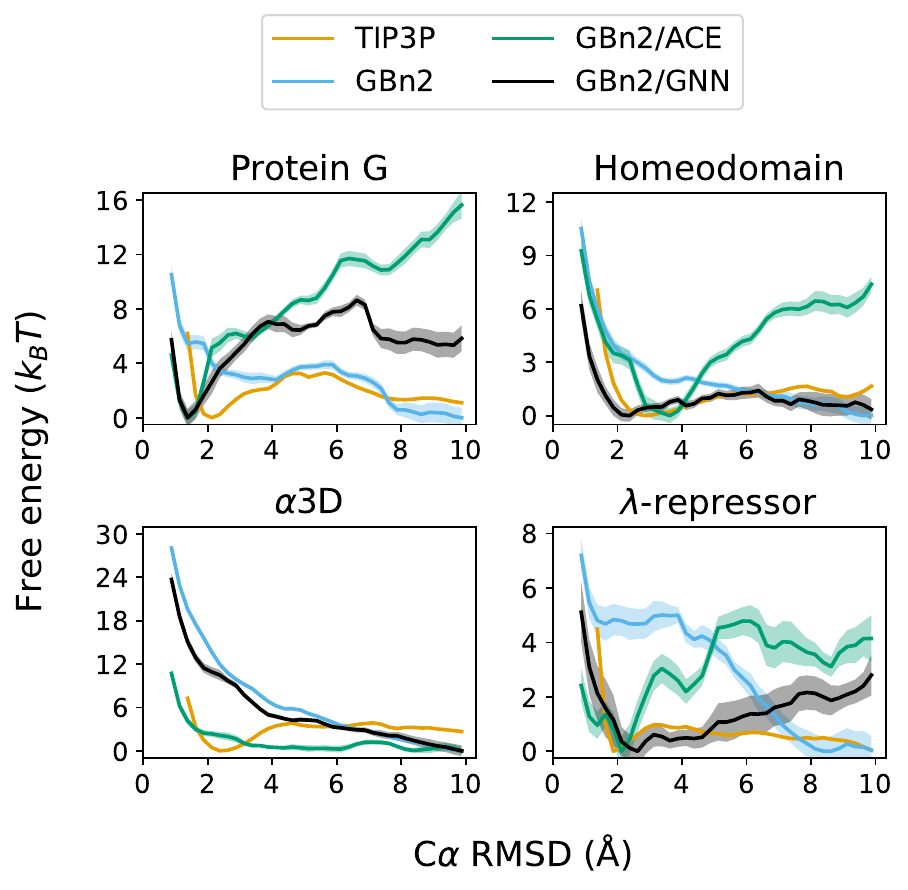}
\caption{
\textbf{The Schake GNN energy, when combined with GBn2,  accurately reproduces protein folding landscapes.} 
Umbrella sampling simulations were performed using three different ISMs, and TIP3P results were obtained from \citet{lindorff-larsen_how_2011}. 
}
\label{free_energy}
\end{figure}

To evaluate the ability of the multi-state formulation to reproduce protein folding landscape, we performed 400~ns umbrella sampling simulations for the four largest fast-folding proteins from \citet{lindorff-larsen_how_2011}. Because $E_\mathrm{GNN}^{\mathrm{ms}}$ is a local motif-based solvation term and does not explicitly encode long-range electrostatic screening, we incorporated it as a correction to the GBn2 ISM, which supplies the continuum electrostatics absent from the GNN. The incorporation of GBn2 also introduces additional stabilization of compact, folded conformations, necessitating the use of a smaller scaling factor ($\gamma = 0.175$) than in the standalone GNN simulations of the previous section. Accordingly, we compared three ISMs: GBn2, GBn2 with the ACE nonpolar term (GBn2/ACE), and GBn2 corrected with $E_\mathrm{GNN}^{\mathrm{ms}}$ (GBn2/GNN). The resulting free energy profiles (Fig.~\ref{free_energy}) were benchmarked against TIP3P explicit solvent\cite{jorgensen_comparison_1983}.

For protein G, homeodomain, and $\lambda$-repressor, GBn2/ACE correctly stabilizes the folded state but substantially under-favors unfolded configurations relative to TIP3P. GBn2/GNN remedies this imbalance: it retains the correct folded minimum while more accurately recovering the unfolded-state free energies. For $\alpha$3D, however, all ISMs fail to produce a folded minimum. Increasing $\gamma$ in the GBn2/GNN model restores the folded basin and yields close agreement with TIP3P (Fig.~S9).

A key design principle of the multi-state energy is that it should produce a more balanced landscape than the one-state formulation. As a protein moves away from its native structure, the energy should not simply rise monotonically, as occurs with $E_\mathrm{GNN}^{\text{os}}$, but should instead allow the model to switch to alternative SS8 motifs that are more prevalent in partially folded or unfolded conformations. This switching behavior can lower the energy of non-native structures and is essential for producing realistic free-energy landscapes in which unfolded states may also correspond to local minima.

To examine whether Schake exhibits this behavior, we analyzed its SS8 predictions for homeodomain along the umbrella sampling trajectories, focusing on residues belonging to its three helices. Since GBn2/GNN closely matches the explicit-solvent free energy profile for this protein (Fig.~\ref{free_energy}), it provides an ideal test case. Across all three helices, Schake’s predicted helix and coil probabilities align with the expected motifs in folded structures and, importantly, transition toward coil-like motifs as more unfolded configurations are sampled (Figs.~S10–S12). These results demonstrate that Schake dynamically switches the most probable motif as the local environment changes, confirming that $E_\mathrm{GNN}^{\text{ms}}$ functions as intended and can assign low energies to both folded and unfolded states when appropriate.

We next examined intrinsically disordered proteins, where multi-state behavior is essential. We selected three proteins (IDP~9, 12, 30) from a recently reported explicit solvent simulation dataset \cite{wang_sequence-dependent_2025} and performed four 100~ns ML/MD simulations at a slightly elevated temperature (350~K) to enhance conformational sampling. Simulations were performed under vacuum, GBn2, GBn2/ACE, and GBn2/GNN ($\gamma=0.175$). Despite the higher temperature, vacuum and GBn2/ACE still collapse the chains into compact structures, deviating substantially from the extended ensembles observed in explicit solvent; this is a known limitation of existing ISMs for modeling IDPs. In contrast, GBn2/GNN produces extended configurations consistent with the TIP3P reference. Although the GBn2/GNN ensembles are slightly more expanded, the model nonetheless places SS8 motifs in the correct regions (Figs.~S13--S15). 
Notably, GBn2 without any additional correction term most-closely matches the TIP3P reference. However, it is still significant that GBn2/GNN does not collapse the chains, especially since our training set predominantly contained folded structures. These results indicate that the multi-state formulation substantially improves IDP modeling relative to existing ISMs, with further tuning likely to yield quantitative accuracy.

\begin{figure}[t!]
\centering
\includegraphics[width=\textwidth]{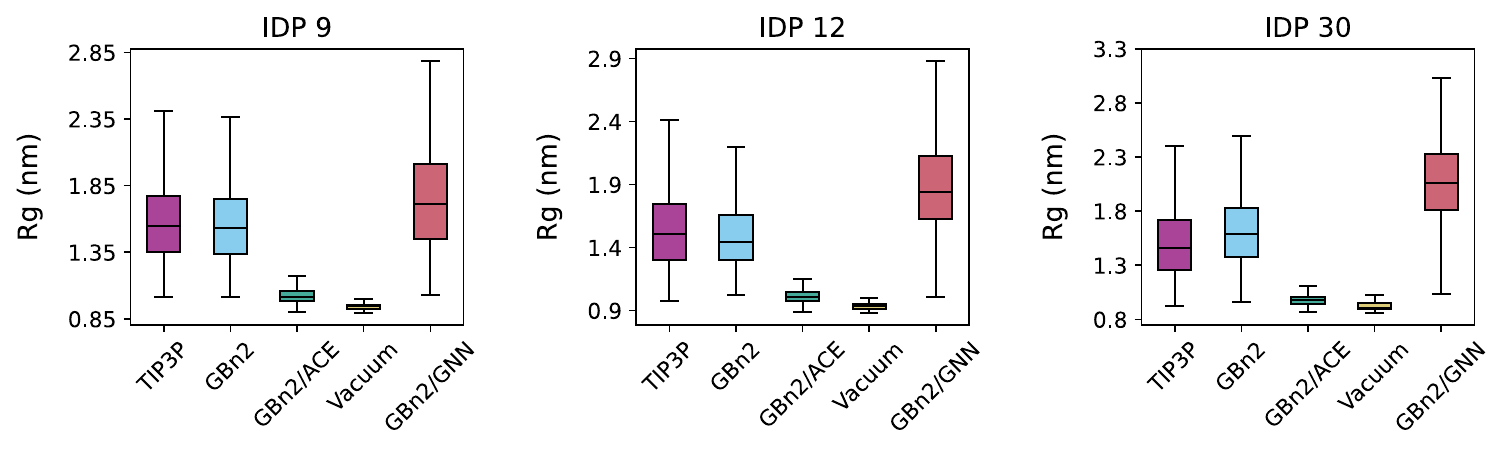}
\caption{
\textbf{ML/MD simulations prevent the collapse of disordered proteins.} 
The solid bar inside each box represents the median radius of gyration value for the three IDPs under different solvent treatments. The bottom and top boundaries of each box represent the first and third quartile values, respectively. The bottom and top whiskers encompass values that are 1.5 times the inter-quartile range below the first quartile and above the third quartile, respectively.
TIP3P reference values were obtained from \citet{wang_sequence-dependent_2025}. 
}
\label{helix_idps}
\end{figure}

\section{Conclusions and Discussion}

In this work, we demonstrate that a data-efficient, multiscale GNN, Schake, can successfully distill secondary-structure distributions learned by the multimodal protein language model ESM3. Schake reproduces SS8 motif likelihoods with high fidelity and generalizes effectively to proteins larger and more diverse than those seen during training. This result alone underscores the surprising efficiency with which evolutionary statistics encoded in protein language models can be transferred to compact GNN architectures.

Building on this representation, we show that the learned GNN potential can function directly as a molecular mechanics energy: it reliably differentiates folded from unfolded configurations, assigns the folded state as energetically favorable, and supports long-timescale ML/MD simulations that preserve key structural features. Across 6.8 $\mu$s of simulations on 11 proteins, Schake maintains near-native conformations and exhibits correlated energetic and structural fluctuations characteristic of physically meaningful folding landscapes.

To extend applicability beyond folded proteins, we introduced a multi-state energy formulation that corrects the GBn2 implicit solvation electrostatics using the GNN-derived potential. This hybrid model accurately reproduces folding free-energy profiles for fast-folding proteins and, notably, captures secondary-structure propensities of intrinsically disordered proteins. The ability to model both folded and disordered states with a single, efficient potential provides strong evidence that distilled protein-language-model statistics offer a powerful basis for implicit solvation modeling.

While these results establish a compelling proof of principle, our present model is not yet a production-ready ISM. Schake was trained primarily on folded structures, and future efforts will benefit from expanding the training set to better represent IDPs, systematically fine-tuning against explicit-solvent simulations, and optimizing GPU kernels to unlock even greater simulation throughput. Nevertheless, the success of our approach demonstrates that knowledge distillation from protein language models is a viable route toward scalable, transferable ISMs. We anticipate that this strategy will enable the development of next-generation ISMs capable of robustly capturing the thermodynamics of proteins across the folded–disordered spectrum.

\section{Methods}
\subsection*{Training Schake to predict solvent-sensitive SS8 motifs}

\subsubsection*{Secondary structure targets from DISPEF and ESM3}
To train Schake to predict solvent-sensitive secondary structure motifs from protein conformations, we used the DISPEF dataset introduced previously.\cite{airas_scaling_2025}
We focus on the DISPEF-M subset, which contains 19,200 training proteins and 4,800 testing proteins, spanning sizes from 16 to 399 amino acids.
For each protein, we extracted Cartesian coordinates for the backbone C$\alpha$, carbonyl carbon (C), and amide nitrogen (N) atoms, along with the corresponding amino acid identities.

Secondary structure (SS8) motif labels were assigned using both model predictions and reference annotations.
For each protein sequence, SS8 likelihoods were predicted using the ESM3-open protein language model. Version 3.2.0 of the \texttt{esm} package was used.
Amino acid sequences were first tokenized using the \texttt{EsmSequenceTokenizer}, and the resulting token sequence $\bm{t}$ was passed through the forward function of ESM3-open.
For the $k$th protein containing $n_\mathrm{res}$ residues, the model outputs a matrix of SS8 logits (``pseudo-energies'') $\bm{S}^{\mathrm{ESM3}}_k \in \mathbb{R}^{n_\mathrm{res} \times 8}$.
These logits were converted to SS8 likelihoods for each residue $i$ via a softmax transformation,
\begin{equation}
p_i^{(j)}(\bm{t}) = 
\frac{\exp\!\left[s_i^{(j)}(\bm{t})\right]}
{\sum_{j \in S_i} \exp\!\left[s_i^{(j)}(\bm{t})\right]},
\tag{3}
\end{equation}
yielding a probability matrix $\bm{P}^{\mathrm{ESM3}}_k \in \mathbb{R}^{n_\mathrm{res} \times 8}$.

Each column corresponds to one of the eight DSSP secondary structure classes, ordered as follows:
$3_{10}$ helices, $\alpha$ helices, $\pi$ helices, hydrogen-bonded turns, $\beta$ sheets, $\beta$ bridges, non-hydrogen-bonded bends, and all remaining structures.

Reference SS8 labels were computed directly from atomic coordinates using the DSSP algorithm as implemented in the \texttt{mdtraj} package (version 1.10.1).\cite{kabsch_dictionary_1983}
These labels were encoded as one-hot matrices $\bm{Y}^{\mathrm{DSSP}}_k \in \mathbb{R}^{n_\mathrm{res} \times 8}$.

Because Schake operates on backbone atoms rather than residues, each residue-level SS8 probability or label was triplicated for the corresponding C$\alpha$, C, and N atoms.
As a result, both $\bm{P}^{\mathrm{ESM3}}_k$ and $\bm{Y}^{\mathrm{DSSP}}_k$ were expanded to dimension $(3n_\mathrm{res} \times 8)$ for ESM3 inference and GNN training.

\subsubsection*{Schake architecture for SS8 likelihood prediction}

The Schake GNN architecture is described in detail in prior works.\cite{schutt_schnet_2018,wang_spatial_2023,airas_scaling_2025}
Originally developed to predict scalar energies for all-atom protein structures, Schake was modified here to predict SS8 motif likelihoods for backbone-only representations.
Each backbone atom is assigned an initial embedding
\begin{equation}
\bm{h}^0 = \left[ \bm{h}_\mathrm{s}, \bm{h}_\mathrm{names} \right],
\tag{4}
\end{equation}
where $\bm{h}_\mathrm{s}$ and $\bm{h}_\mathrm{names}$ are trainable vectors (16 dimensions each) associated with the amino acid identity and backbone atom type, respectively. These embeddings are then passed through two message passing layers to produce updated embeddings $\bm{h}^L$.
Given an input configuration $\bm{x}$ for protein $k$, Schake outputs a matrix of SS8 pseudo-energies
$\bm{S}^{\mathrm{GNN}}_k \in \mathbb{R}^{3n_\mathrm{res} \times 8}$.
These outputs are obtained by passing the updated embeddings $\bm{h}^L$ through a three-layer feed-forward network.
All architectural hyperparameters are reported in Tab.~S4.

For backbone atom $i$, the predicted likelihood of SS8 motif $j$ is obtained via a softmax,
\begin{equation}
q_i^{(j)}(\bm{x}) = 
\frac{\exp\!\left[s_i^{(j)}(\bm{x})\right]}
{\sum_{j \in S_i} \exp\!\left[s_i^{(j)}(\bm{x})\right]}.
\tag{5}
\end{equation}
These likelihoods define a local motif energy
\begin{equation}
E_i^{(j)}(\bm{x}) = -\log\!\left(q_i^{(j)}(\bm{x}) + \epsilon\right),
\tag{6}
\end{equation}
which contributes to the total GNN-derived energy as defined in Eqs.~\ref{Eq:01} and~\ref{new_e}. To ensure numerical stability, $\epsilon = 10^{-8}$ was added.

\subsubsection*{Knowledge distillation and training procedure}

Schake was trained using a knowledge distillation framework,\cite{hinton_distilling_2015} with ESM3-open serving as the teacher model.
Training was performed using PyTorch (version 2.0)\cite{paszke_pytorch_2019} and PyTorch Geometric (version 2.3.1).\cite{fey_fast_2019}

The primary loss was an averaged cross-entropy loss $\ell_\mathrm{CEL}$ (Eq.~S2) between the ESM3-predicted SS8 likelihoods $\bm{P}^{\mathrm{ESM3}}$ and the Schake-predicted likelihoods $\bm{Q}^{\mathrm{GNN}}$.
To further anchor predictions to physical structure, a second cross-entropy term between the DSSP-derived one-hot labels $\bm{Y}^{\mathrm{DSSP}}$ and $\bm{Q}^{\mathrm{GNN}}$ was included.
The total training loss was therefore
\begin{equation}
\ell_\mathrm{tot} =
\ell_\mathrm{CEL}\!\left(
\{\bm{P}^{\mathrm{ESM3}}_k\}_{k=1}^N,
\{\bm{Q}^{\mathrm{GNN}}_k\}_{k=1}^N
\right)
+
\ell_\mathrm{CEL}\!\left(
\{\bm{Y}^{\mathrm{DSSP}}_k\}_{k=1}^N,
\{\bm{Q}^{\mathrm{GNN}}_k\}_{k=1}^N
\right).
\label{cel_loss_tot}
\tag{7}
\end{equation}

Training and test-set accuracies are reported in Fig.~\ref{confusion_mats} and Fig.~S1, respectively.
Models were trained for 120 epochs using a batch size of 50 proteins.
Optimization was performed with the Adam optimizer,\cite{kingma_adam_2017} using an initial learning rate of $10^{-3}$, decayed by a factor of 0.9 every three epochs.
L2 regularization,
$\ell_\mathrm{reg} = \lambda_\mathrm{reg} \sum_i \theta_i^2$ with $\lambda_\mathrm{reg} = 10^{-6}$,
was applied via weight decay during optimizer initialization.

\subsection*{Classical MD and ML/MD simulation details}

All classical MD and ML/MD simulations were performed using OpenMM (version 8.0.0).\cite{eastman_openmm_2024}
ML/MD simulations employed the OpenMM-Torch plugin (version 1.0) with PyTorch (version 1.11.0), PyTorch-Cluster (version 1.5.9), and PyTorch-Scatter (version 2.0.8).\cite{paszke_pytorch_2019,fey_fast_2019}

Initial structures for CspA, Top7, Hyp Protein, and NTL9$_\mathrm{52}$ were obtained from the Protein Data Bank\cite{berman_protein_2000} (structure IDs 1MJC,\cite{schindelin_crystal_1994} 1QYS,\cite{kuhlman_design_2003} 1WHZ, and 2HBA,\cite{cho_energetically_2014} respectively), while the initial structures for the proteins shown in Figs.~\ref{free_energy} and \ref{helix_idps} were obtained from previous works.\cite{lindorff-larsen_how_2011,wang_sequence-dependent_2025}
These initial structures also serve as the reference structures for all RMSD calculation.
Simulation inputs were prepared using \texttt{tleap} from AmberTools23.\cite{case_ambertools_2023}

Unless otherwise noted, simulations used the ff14SBonlysc vacuum force field.\cite{hornak_comparison_2006,maier_ff14sb_2015}
Unfolding simulations presented in Fig.~S4 were performed using the CHARMM36 vacuum force field.\cite{best_optimization_2012}
Additional simulation parameters and integration details are provided in the \emph{Supporting Information},\cite{zhang_unified_2019,hopkins_long-time-step_2015}
and performance benchmarks are summarized in Tabs.~S5--S7.
Trajectory analysis was performed using \texttt{mdtraj} (version 1.10.3)\cite{mcgibbon_mdtraj_2015}
and ChimeraX (version 1.10.1)\cite{pettersen_ucsf_2021}.
Umbrella sampling free energy profiles (Fig.~\ref{free_energy}) were computed using the FastMBAR package (version 1.4.3).\cite{ding_fast_2019}

\section{Supporting Information}

    Expanded equations, simulation setup information, and additional supporting figures and tables.

\section{Data Avaiability}

    Code for our modified Schake architecture will be made available on GitHub at the following link: \url{https://github.com/ZhangGroup-MITChemistry/Schake_GNN/}. We also will include examples of how to use Schake in ML/MD simulations. Additionally, the optimized model parameters from training will be provided.

\begin{acknowledgement}

    This work was supported by the U.S. National Institutes of Health (Grant R35GM133580). 
    J.A. was supported by the the U.S. National Science Foundation Graduate Research Fellowship. We thank the MIT Office of Research Computing and Data for providing high performance computing resources.
    This work also used Bridges-2 at the Pittsburgh Supercomputing Center through allocations BIO240299 and CHE240139
    from the Advanced Cyberinfrastructure Coordination Ecosystem: Services \& Support (ACCESS) program, which is supported by U.S. National Science Foundation grants \#2138259, \#2138286, \#2138307, \#2137603, and \#2138296.

\noindent\textbf{Competing interests:} 
Authors declare that they have no competing interests.

\end{acknowledgement}

% \begin{suppinfo}

% This will usually read something like: ``Experimental procedures and
% characterization data for all new compounds. The class will
% automatically add a sentence pointing to the information on-line:

% \end{suppinfo}


\clearpage
\newpage
\bibliography{main}

\end{document}